\documentclass[12pt, preprint]{aastex}
\usepackage{amsmath}

\begin{document}

\title{Near-ultraviolet and optical effects of debris disks around White Dwarfs}

\author{A. Zabot\altaffilmark{1} and A. Kanaan\altaffilmark{1} and
        R. Cid Fernandes\altaffilmark{1}}
\affil{Departamento de F\'isica, Universidade Federal de Santa Catarina,
    Florian\'opolis, CP 476, 88010-970, Brazil}

\altaffiltext{1}{Departamento de F\'isica, Universidade Federal de Santa
                 Catarina, Florian\'opolis, CP 476, 88010-970, Brazil}

\begin{abstract}
   Studies of debris disks around white dwarfs (WDs) have focused on
   infrared wavelengths because debris disks are much colder than the
   star and are believed to contribute to the spectrum only at longer
   wavelengths. Nevertheless these disks are made of dust grains which
   absorb and scatter near-UV and optical photons from the WD, leaving a
   fingerprint that can be used to further constrain disk
   properties. Our goal is to show that it is
   possible to detect near-UV and optical effects of debris disks in
   the star + disk integrated spectrum. We make theoretical
   calculations and discuss the necessary observational conditions to
   detect the near-UV and optical effects. We show how these effects
   can be used to infer the disk mass, composition, optical depth, and
   inclination relative to the line of sight.  If the IR excess is due
   to a disk, then near-UV and optical effects should be observed in
   only some systems, not all of them, while for dust shells the
   effects should be observed in all systems.

\end{abstract}


\section{Introduction}

White dwarfs (WDs) are degenerate stellar nuclei with a mass roughly that of the
Sun and radii one hundredth that of the Sun; consequently,
their surface gravity is $\sim$$10^4$ greater than the
Sun's. \cite{lie84} identified some odd WDs with metal-rich
atmospheres. With such a powerful gravity pulling the chemical
elements toward the stellar nucleus, it is somewhat unusual to have a metal-rich
atmosphere. In fact, the timescales for an element heavier than
hydrogen or helium to sink are small: $\sim$$10^2$ yr in WDs with
hydrogen atmospheres (DAs) and $\sim$$10^5$ yr with helium atmospheres
(DBs) \citep{jur08,von07,paq86}.

Interstellar material accretion onto the WD surface was one of the first
explanations for the metal-rich atmospheres. Knowing the diffusion
timescales of metals in the stellar atmospheres and the metallicity of a given
star, it is possible to calculate the necessary accretion rate to keep this
metallicity constant in time \citep{koe06}. Typical values are
$10^{-18}$ to $10^{-15}$ M$_\odot$yr$^{-1}$. These values are too high to be
explained exclusively by interstellar accretion. Furthermore, if there were
accretion from the interstellar medium onto the DBs, there should be a large
amount of hydrogen pollution in their spectra, but this pollution has not been
detected \citep{dup1993,far2008}.

\cite{zuc87} observed an IR~excess in the spectrum of G29-38. The shape of this
IR~excess is a bump which peaks at $\sim$$10~\mu m$. Its width is roughly
$20~\mu m$ and can be fitted with a blackbody of $T_{\mathrm{eff}} \sim 10^3$ K.
\cite{gra90} argued that an asteroid closely approaching G29-38 could explain
this infrared excess. When the asteroid orbit reaches the Roche radius it is
disrupted and forms a disk around the star. This disk is heated up by the
stellar radiation and emits in the infrared. The disk material falls down
continuously on the WD giving rise to the observed metal-rich atmospheres.

Using a disk model, Jura \& collaborators (2003,2007,2008) were able to fit
the IR excess of many WDs. Through these fits they determined disk physical
parameters. With a different approach, \cite{rea05} fit the spectrum to a thin
dust shell model.

While searching for interacting binary WDs, 
\cite{gan2006,gan2007,gan2008} observed double peaks in calcium lines
in a few DAZs and DBZs.  Although these observations strongly suggest
the disk hypothesis,it is still possible that in some cases the emitting
region could be a torus or a shell \citep{rea09} rather than a disk.

Previous works focused on the IR emission properties of
debris disks around the WD. In this work, we propose a new and complementary
observational test looking at the absorption and scattering properties instead.
We develop a simple theoretical framework to predict what will be observable
and measurable according to the properties of the system. We also suggest an
observational program to reach our goal.

We investigate the possibility of detection of debris disks effects
in the near-UV and optical. We start the analysis in the limit of an optically
thick disk in Section~\ref{sec:od} and in Section~\ref{sec:ot} we extend this to
the optically thin limit case. In Section~\ref{ObsTest}, we discuss the
observational predictions of our models. Our conclusions are presented in
Section~\ref{sec:conc}.

\section{Opaque disk}
\label{sec:od}

  The optically thick limit is the natural first approach to investigate the
possible effects of a debris disk in the spectrum of a WD. This limit can be
achieved not only in massive disks but also in certain regions of all disks
specially if the disk has some gas like some recently discovered gaseous disks
\citep{gan2006,gan2007,gan2008}. Also, the mathematical treatment developed in
this section will be used in the next section when dealing with the
optically thin limit.

A completely opaque disk will not have any
spectral features because it is totally opaque and absorbs any photon whatever
its energy. The only effect of an obscuring disk will be a decrease in the
received flux from the star. The most the disk can obscure the star is half of
the projected stellar surface, $\pi R_{wd}^2/2$. The increase in the apparent
magnitude of the star will be $0.75$ mag.

This value is much higher than current photometric accuracy, and if present,
would have been detected for those WDs which have an observed parallax, a good
spectrum and an IR~excess. The ``good spectrum' permits a determination of
$T_{\mathrm{eff}}$ and $\log g$ and hence a luminosity. Clearly the luminosity
inferred  from the parallax should agree with the luminosity inferred from the
spectral fit. At the very least, this procedure will allow us to affirm that
there is not a big opaque disk in the known WDs with IR~excess or, at least,
this disk is not in a favorable inclination.

  For the general case of a disk with arbitrary inclination and any combination
of inner and outer radii, the flux received at the Earth from the system
(target) is,

\begin{equation}
  \begin{aligned}
  F^{target} &=\lefteqn { \int_{\Omega_{wd}} I \cos \theta d\Omega }\\
     &= I \left(\frac{R_*}{D}\right)^2
      \int_{0}^{\pi/2} \int_{\phi_{min}}^{\phi_{max}}
      \sin \theta \cos \theta d \phi d \theta 
  \end{aligned}
\end{equation}

\noindent
where we assumed that the intensity (I) is uniform over the stellar surface.
The stellar radius is $R_*$ and $D$ is the distance from the Earth to the system.
There are three possible projections for the disk, as seen in Figure~\ref{3disks}.
This gives different values of $\phi_{min}$ and $\phi_{max}$:

\begin{figure}
\epsscale{0.6}
\plotone{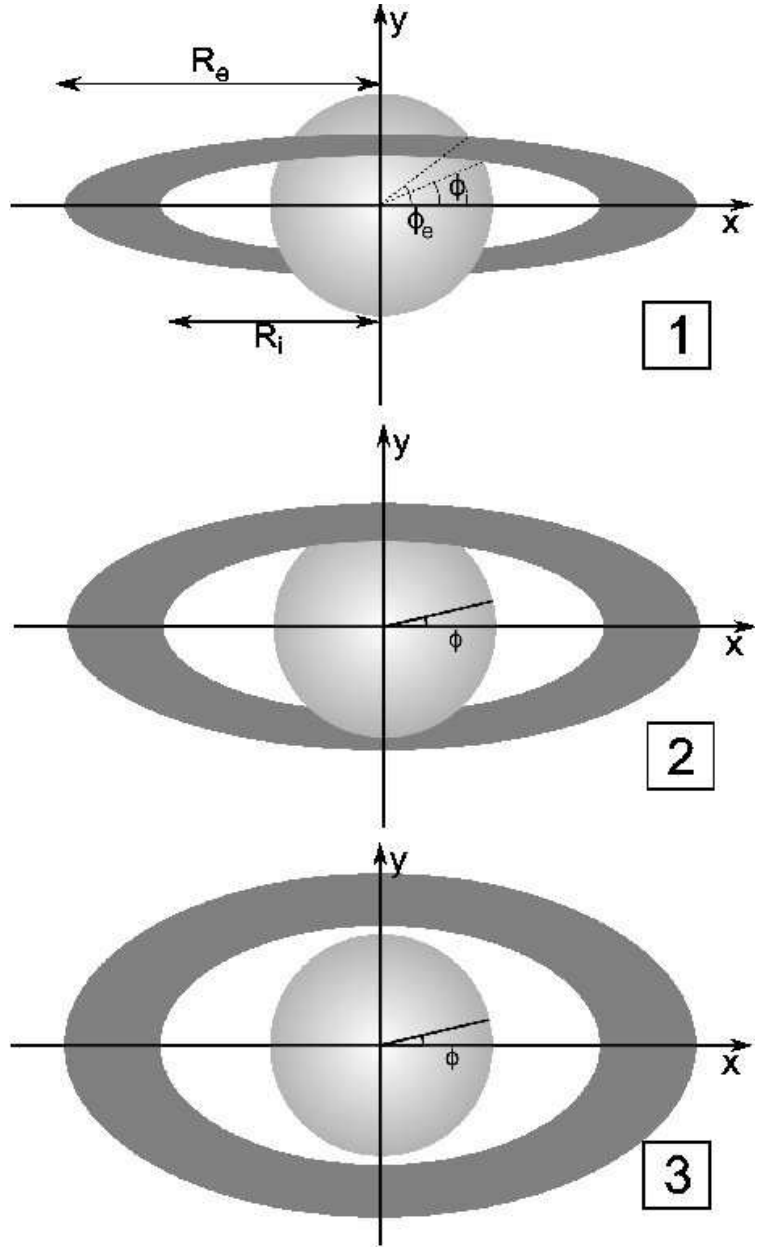}
\caption{Debris disk and WD. For a given size and inclination of the disk,
         different amounts of it obscure the star. The $x-$ and $y-$axes are in
         the sky-plane and the observer is over the $z-$axis which makes an
         angle $i$ with the normal of the disk. The
         angles $\phi_i$ and $\phi_e$ show where the disk starts and stops
         obscuring the star.
}
\label{3disks}
\end{figure}

  \begin{itemize}
  \item 
  $\int_{\phi_{min}}^{\phi_{max}} d \phi =
      2 \left[ \int_{-\pi/2}^{\phi_i} d \phi + \int_{\phi_e}^{\pi/2} d \phi
      \right ]
      \\= 2\pi - 2( \phi_e-\phi_i )
  $,

  \item
  $\int_{\phi_{min}}^{\phi_{max}} d \phi =
      2 \int_{-\pi/2}^{\phi_i} d \phi    =
      \pi + 2\phi_i
  $,

  \item
  $\int_{\phi_{min}}^{\phi_{max}} d \phi =
      2 \int_{-\pi/2}^{\pi/2} d \phi     =
      2 \pi
  $.

  \end{itemize}

  Using the dimensionless radii $r_{\{i/e\}} \equiv R_{\{i/e\}}/R_{star}$ we
define the function $g \equiv g(\theta,r_i,r_e)$:

  \begin{equation}
  g = \left\{
   \begin{array}{l l}
      \pi - (\phi_e-\phi_i) & \mbox{, $r_e \cos i < \sin \theta $}\\
      \pi/2 + \phi_i        & \mbox{, $r_i \cos i < \sin \theta \leq r_e
                                                             \cos i $}\\
      \pi                   & \mbox{, $r_i \cos i \geq \sin \theta$}\\
    \end{array}
    \right.
  \label{eqg}
  \end{equation}

\noindent
and write the flux:

  \begin{equation}
  F^{target} = I \left(\frac{R_*}{D}\right)^2
            \int_{0}^{\pi/2} \sin(2\theta) g(\theta,r_i,r_e) d \theta
  \label{eqF}
  \end{equation}

\noindent
where

  \begin{equation}
  \phi_{i,e} = \arctan \left[ \cos i \arccos\left(
               \frac{\sqrt{ \sin^2\theta/r_{i,e}^2 - \cos^2 i}}{\sin i}
               \right) \right ]
  \label{eqphi}
  \end{equation}

  The total flux received from an unobscured system is $\pi I
  (R_*/D)^2$.  We call the hypothetical unobscured star ``template''
  and the obscured star ``target''. Defining $p$ as the ratio of the
  obscured to the total projected area:

\begin{equation}
 p \equiv \frac{A_{obscured}}{A_{total}}  = \frac{A_{target}}{A_{template}}
\label{def:p}
\end{equation}

\noindent
we write the increase in magnitude as:

\begin{equation}
 \Delta m = -2.5 \log(1-p)
\label{eq:mp}
\end{equation}

In the completely opaque hypothesis we may obviously write that,

\begin{equation}
  \begin{aligned}
 p &=\lefteqn { \frac{F^{target}}{F^{template}}
   = \frac{F^{target}}{\pi I} \left(\frac{D}{R_*}\right)^2 }\\
   &= \frac{1}{\pi} \int_{0}^{\pi/2} \sin(2\theta) g(\theta,r_i,r_e) d \theta
  \end{aligned}
\label{eq:p}
\end{equation}

\begin{figure}
\epsscale{1}
\plotone{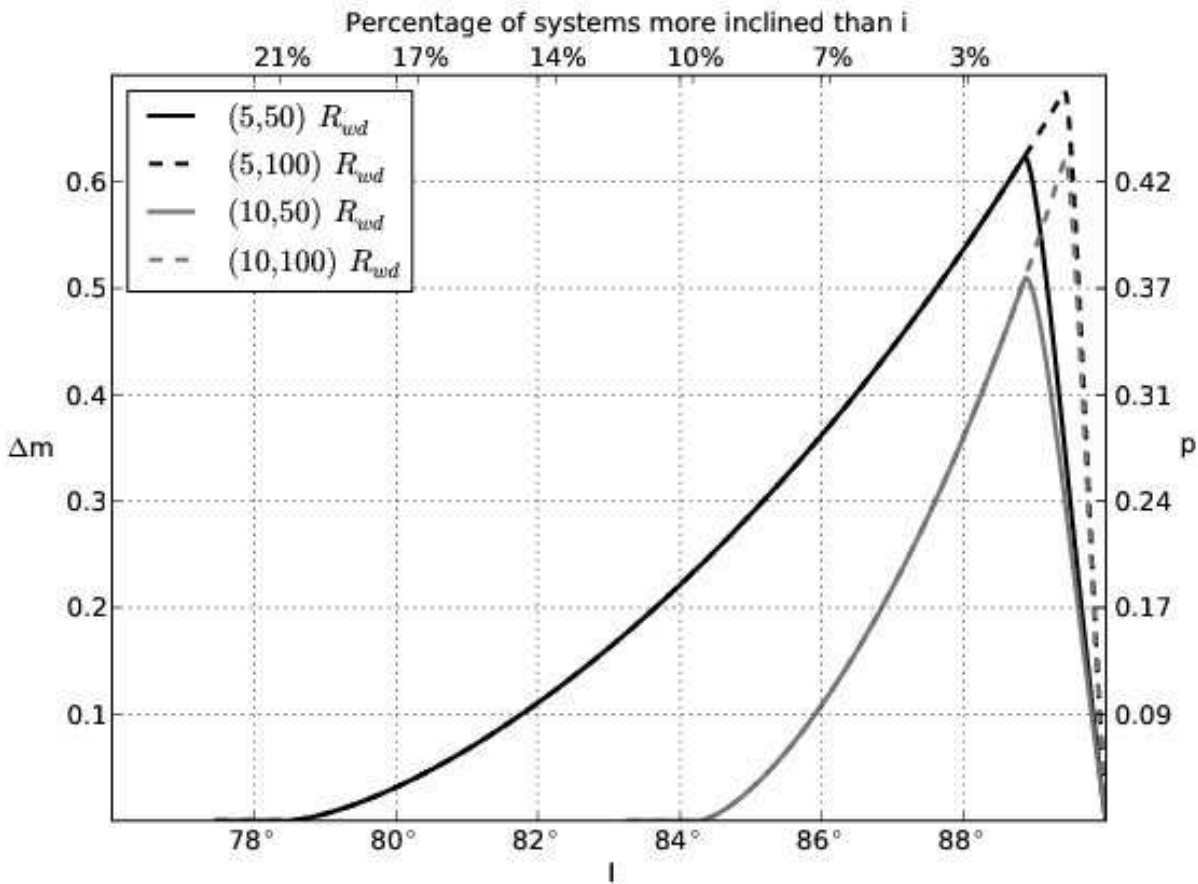}
\caption{Increase in magnitude vs. inclination for opaque disks for
   different inner ($r_i$) and outer ($r_e$) disk radius combinations.
   Systems seen face on have $i=0^\circ$. The fast decrease of $\Delta m$ for
   $i \rightarrow 90^\circ$ is an artifact of the mathematical model that
   assumes an infinitely thin disk. For real disks, there would be a plateau
   lower than curve peak, but the disk should be really thin indeed, so the
   plateau is very close to $\Delta m = 0$ and it actually is not a plateau,
   but rather a single point.
   The top axis label shows the percentage of systems more inclined than $i$.
   The right axis label shows $p$, the ratio of the obscured to the total
   projected area (Equation~\ref{def:p}).}
\label{fig:hidden}
\end{figure}

  The solution of Equation \ref{eq:p} with Equations \ref{eqg} and \ref{eqphi}
gives the flux received from the system, as can been seen in
Figure~\ref{fig:hidden}. The probability of finding a system more inclined than
a given angle is the ratio of the solid angle occupied by these systems
to the total solid angle: $P(i>i_0) = \cos i_0$.

  Figure~\ref{fig:hidden} shows that it is possible to detect a completely
opaque debris disk. For an inclination angle causing any blocking, the bigger
the disk is, the easier it is to detect the effect. The inner and outer radii
are based on physical constrains. If the inner radius is small, the dust
particles sublimate because the
temperature exceeds $\sim 1200$K~\citep{jur07a}. On the other hand, if the
outer edge of the disk is big ($\gtrsim 100 R_{wd}$), the dust grains will be
cold and any emission will be undetectable in practice. The exact
values depend on the dust type and grain size.

  The presence of an obscuring disk might be inferred by comparing the expected
increase in stellar magnitude with the luminosity derived from
$T_{\mathrm{eff}}$ and $\log g$ and parallax. If the observed and the expected
magnitudes are correct and not equal, the flux deficiency can probably be
explained by obscuration from a debris disk.

  We use the measured parallax of GD~362 to illustrate the previous analysis
with one real case. \cite{kil08} obtained d~$= 50.6_{-3.1}^{+3.5}$~pc for
GD~362. Using simple error propagation, we have a rough estimate for the highest
acceptable difference between expected and measured magnitude: $\sigma_m
\approx 0.15$ mag.  \cite{kil08} did not find any discrepancies between parallax
and flux, implying no obscuration of the star.

From a geometrical perspective this is expected since from
Figure~\ref{fig:hidden} $\Delta m\approx0.15$ mag implies an inclination higher
than $\sim$$80^\circ$ and less than $\sim$$20\%$ of the systems will be more
inclined than this. Indeed, \cite{jur07b} showed that GD~362 must be seen
nearly face on to be able to reproduce its IR~excess flux with physically
reasonable inner and outer radii. Assuming an almost edge on system would
require a big disk and an unusual mechanism to heat it to reproduce the
measured IR~excess flux. Therefore, our work is in agreement with the previous
results and this analysis illustrates what kind of study must be done with
other systems which may be found to be nearly edge on.

\section{Optically thin dust disk}
\label{sec:ot}

  After having derived the basic concepts of the problem with the optically
  thick limit we generalize the equations to the optically thin limit.
  The ratio ($\xi_\nu$) of the flux from obscured (target) to the equivalent
star with no obscuration (template) is composed of three main components:

\begin{equation}
\begin{aligned}
   \xi_\nu &\equiv \lefteqn {\frac{F^{target}_\nu}{F^{template}_\nu}}\\ &=
   \left. \xi_\nu \right|_{unobscured} +
   \left. \xi_\nu \right|_{obscured} +
   \left. \xi_\nu \right|_{scattered}
\end{aligned}
\label{eq:3comp}
\end{equation}

  The unobscured ratio component is simply $1-p$. The obscured ratio
  is given by $p e^{-\tau_\nu^{ext} /\cos{i}}$. The extinction optical depth
  $\tau_\nu^{ext}$ of the disk regions obscuring the star accounts for the
  absorbed and scattered light along the line of sight to the star.
  The scattered component comes from the disk regions which do not obscure the
  star but scatter photons to the line of sight. There is no emission component
  because the dust temperature is lower than the dust sublimation temperature
  ($\sim$1200~K) and thus the dust emission only contributes in the infrared.

  Using the dimensionless extinction efficiency ($Q^{ext}_\nu$) instead of
extinction cross section ($[C^{ext}_\nu] = cm^2$) we write the differential
extinction optical depth in the disk as:

\begin{equation}
 d\tau_\nu = n C^{ext}_\nu dz = n Q^{ext}_\nu \pi a^2 dz,
\end{equation}

\noindent
where $n$ (cm$^{-3}$) is the number of dust grains per unit volume, $a$ (cm)
is the grain radius and $z$ (cm) is the vertical dimension of the disk

 The disk volume density ($\rho$) is related to the density of a typical dust
grain ($\rho_d$) through,

\begin{equation}
 \rho = \frac{4}{3} \pi a^3 \rho_d n .
\label{rho}
\end{equation}

  Assuming the disk to be vertically uniform we integrate to write

\begin{equation}
 \tau_\nu^{ext} = \int^{H/2}_{-H/2} \frac{3 Q^{ext}_\nu \rho}{4 a \rho_d} dz
          = \frac{3 Q^{ext}_\nu \rho}{4 a \rho_d} H
          = \frac{3 Q^{ext}_\nu \Sigma}{4 a \rho_d},
\label{eq:tau}
\end{equation}

\noindent
where $\Sigma$ (g/cm$^2$) is the disk surface density and $H$ is the disk
height.

  We define:

\begin{equation}
 \tau_0 = \frac{3 \Sigma}{4 a \rho_d}
\label{eqt0}
\end{equation}

\noindent
and write Equation~\ref{eq:3comp} as

\begin{equation}
 \xi_\nu = (1-p) +
               p e^{ -\tau_0 Q^{ext}_\nu / \cos{i} } +
               \left. \xi_\nu \right|_{scattered}.
\label{eqFd}
\end{equation}

  To simplify the scattering term, we assume isotropic and
coherent scattering and also that the light is not attenuated before and after
being scattered by the disk. The last hypothesis is valid in the optically thin
case and causes an overestimation of the scattering because we ignore the
absorbed photons. The scattered intensity is given by

\begin{equation}
 I^{sca}_\nu = \epsilon_\nu \frac{H}{\cos{i}}
             = \pi a^2 Q^{sca}_\nu n J^{wd}_\nu \frac{H}{\cos{i}},
\end{equation}

\noindent
where $\epsilon_\nu$ is the emissivity, $Q^{sca}_\nu$ is the scattering
efficiency and $J^{wd}_\nu$ is the mean stellar intensity

\begin{equation}
  J^{wd}_\nu = \frac{I^{wd}_\nu \pi R_{wd}^2}{4 \pi r^2}
\end{equation}

  Ignoring the disk regions hidden by the star we integrate over the disk
surface to get the flux:

\begin{equation}
  F_\nu = \frac{1}{2} \frac{3 \Sigma}{4 a \rho_d} Q^{sca}_\nu
          \ln \left( \frac{re}{ri} \right ) \cos{i}
          \, \pi I^{wd}_\nu \left( \frac{R_{wd}}{D} \right)^2
\end{equation}

\noindent
using Equation~\ref{eqt0} and dividing by the template flux,

\begin{equation}
  \left. \xi_\nu \right|_{scattered} = \frac{1}{2} \tau_0 Q^{sca}_\nu
          \ln \left( \frac{re}{ri} \right ) \cos{i},
\end{equation}

\noindent
which allows us to write Equation~\ref{eqFd} as

\begin{equation}
 \xi_\nu = (1-p) +
               p e^{ -\tau_0 Q^{ext}_\nu / \cos{i} } +
               \frac{1}{2} \tau_0 Q^{sca}_\nu
               \ln \left( \frac{re}{ri} \right ) \cos{i}
\label{eqFdcp}
\end{equation}

Besides the parameters $p$ and $\tau_0$, we have the absorption efficiencies
which are characteristic of the dust type. We used the tables of optical
constants of silicate glasses from \cite{dor95}. The
authors prepared two different glasses in laboratory: pyroxene,
Mg$_x$Fe$_{1-x}$SiO$_3$, with $x$=0.4, 0.5, 0.6, 0.7, 0.8, 0.95, 1.0 and olivine,
Mg$_{2x}$Fe$_{2-2x}$SiO$_4$, with $x$=0.4 and 0.5.

\begin{figure}
\epsscale{1}
\plotone{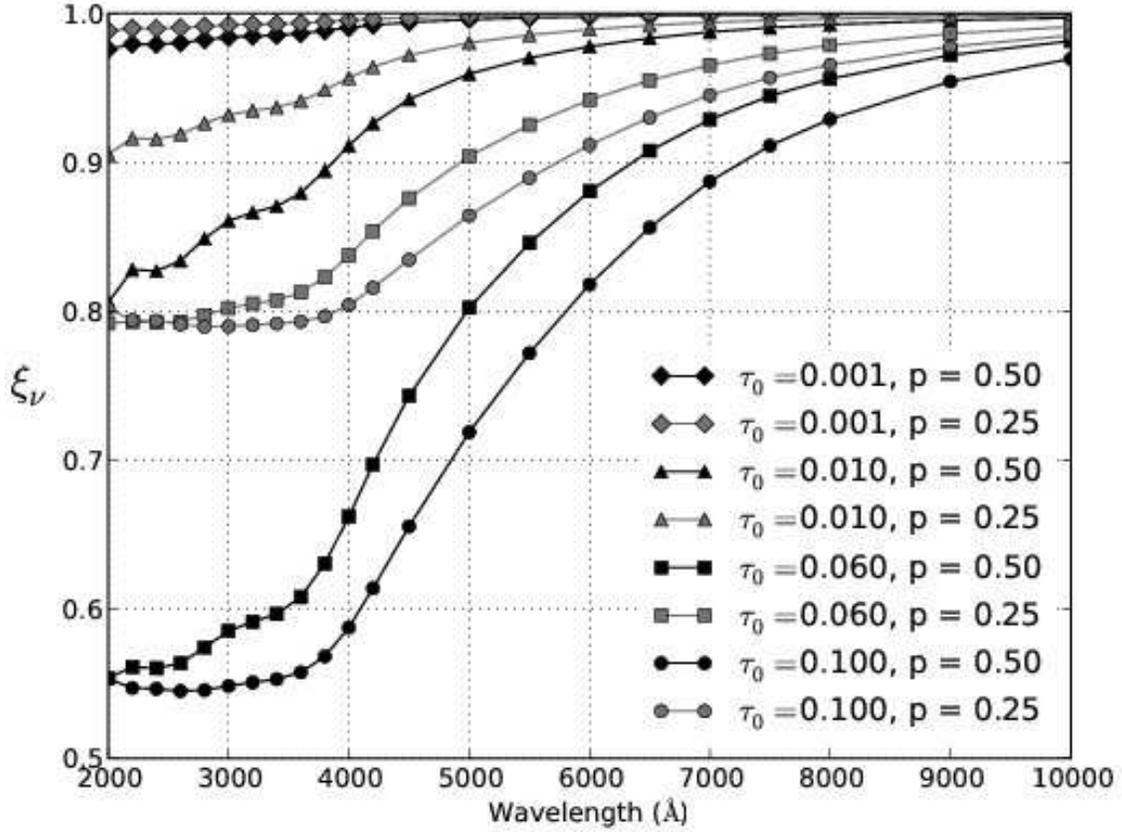}
\caption{Expected ratio between the target and the template star
          (Equation~\ref{eqFdcp}) when the inclination is such that the disk
          obscures
          a part of the WD as in 1 and 2 in Figure \ref{3disks}. The disk is
          composed of olivine Mg$_{0.8}$Fe$_{1.2}$SiO$_4$ dust grains. To
          calculate the scattering we used
          $\ln(r_e/r_i) \lesssim \ln(100/5) = 3$
          as a superior limit in Equation~\ref{eqFdcp} and adopted $i=85^\circ$.}
\label{pyrmg70}
\end{figure}

\begin{figure}
\epsscale{1}
\plotone{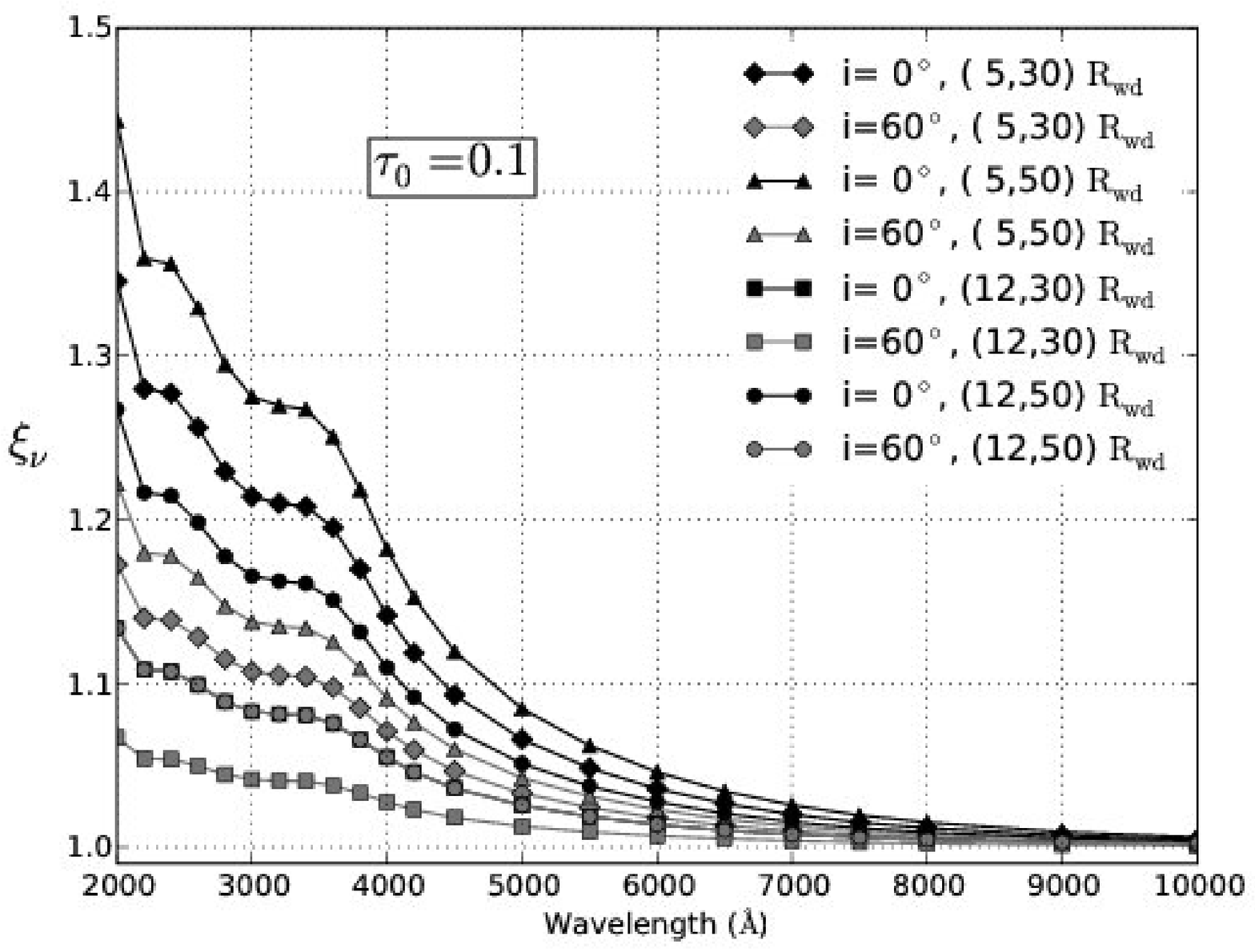}
\caption{Expected ratio between the target and the template star
          (Equation~\ref{eqFdcp}) when the disk does not obscure the WD
          and we see the scattering plus the stellar light. The disk is
          composed of olivine Mg$_{0.8}$Fe$_{1.2}$SiO$_4$ dust grains. All the
          curves were calculated with $\tau_0 = 0.1$ because the pure scattering
          term in Equation~\ref{eqFdcp} is linear in $\tau_0$.}
\label{pyrmg70sca}
\end{figure}

Figures \ref{pyrmg70} and \ref{pyrmg70sca} display the results from
Equation~\ref{eqFdcp} for different optical depths and system
geometries. Figure \ref{pyrmg70} represents inclinations where the disk
obscures the star, and light is absorbed and Figure
\ref{pyrmg70sca} when there is no obscuration and we see only
scattering plus the WD light.  For observational tests, the region
from $\textrm{3000\AA\ to 5000\AA}$ is the most interesting, because it
shows a sharp change in the ratio between the target and the template
which cannot be easily discarded as bad flux calibrations.

\section{Observational test}
\label{ObsTest}

  Our modeling gives rise to direct observational tests. We can test the
  effects of the disk in the near-UV and the optical, dividing the
  spectrum of the target by the template. In the optically thin case, the
  result will be color dependent and can provide physical parameters for
  the disk structure.

  The template star should be as similar to the target star as
  possible. Ideally, it would be the same WD without the obscuring
  disk. As that is not possible to have, we need a similar WD without
  any peculiarity in the spectrum.  Any other WD will differ from the
  target star in $T_{\mathrm{eff}}$ and $\log g$ and this difference
  can make the division of the spectra resemble the expected disk
  effects.

\begin{figure}
\epsscale{1}
\plotone{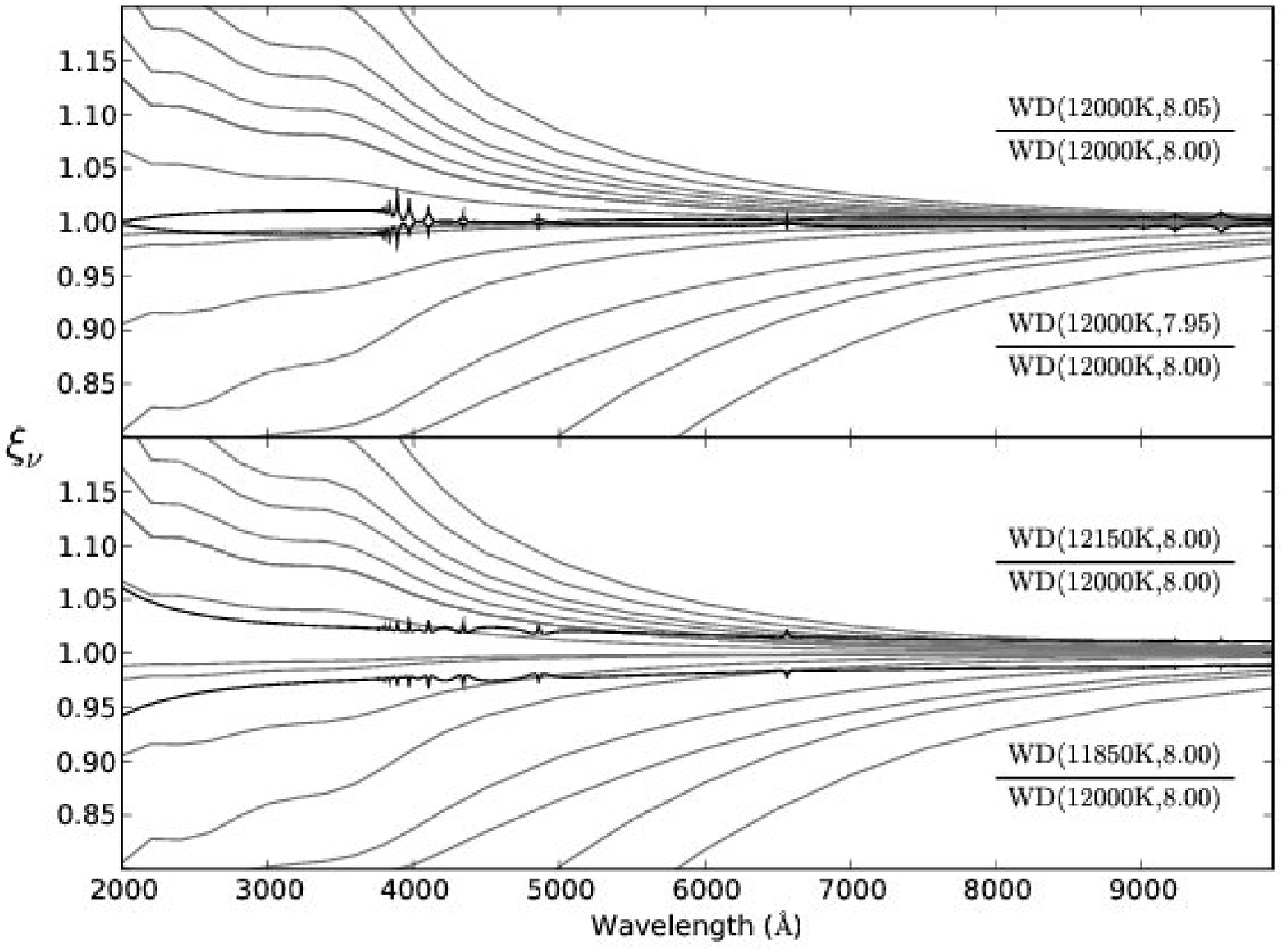}
\caption{Comparison of the effects from debris disk obscuration and the effects
   of small differences in $T_{\mathrm{eff}}$ and $\log g$ between the target and
   the template stars. The gray lines show the expected effects shown in Figures
   \ref{pyrmg70} and \ref{pyrmg70sca}. In the upper panel the black solid
   lines show the ratio of WD spectra with fixed $T_{\mathrm{eff}}$ and
   varying $\log g$. In the lower panel, we fixed $\log g$ and
   varied $T_{\mathrm{eff}}$.}
\label{wdRatio}
\end{figure}

  In Figure~\ref{wdRatio}, we present these effects using theoretical WD
  spectra from \cite{koe08}.  We assume a target star of
  $T_{\mathrm{eff}} = 12,000$~K and $\log g = 8.0$, similar to G29-38
  \citep{rea09}. In the upper panel, we keep $T_{\mathrm{eff}}$ fixed
  and vary $\log g$ by $0.05$~dex and in the lower we keep $\log g$
  fixed and vary $T_{\mathrm{eff}}$ by $150$~K. According to \cite{lie05},
  the uncertainties in temperature are of the order of $1.2 \% \approx 150$~K
  and 0.038 in $\log g$. So, larger temperature or $\log g$ differences would be
  readily noticed. One can see from Figure~\ref{wdRatio} that modification of UV
  flux densities by disks that absorb or scatter light, as in
  Figures \ref{pyrmg70} and \ref{pyrmg70sca}, can be distinguished from
  observational uncertainties in template stars.

  The White Dwarf Catalog \citep{mcc09} currently lists 12,456 stars.
  Therefore it is not too hard to find a template star with a temperature
  similar to the target.  As an example, G29-38 and Ross~548 have
  exactly the same temperature and $\log g$.  One needs to be careful
  about this comparison as these values of $T_{\mathrm{eff}}$ and
  $\log g$ were obtained from different determinations.  When
  comparing the target with template it will be necessary to use
  $T_{\mathrm{eff}}$ and $\log g$ obtained from similar data and the
  same models.

\subsection{Parameter determination}

  It is difficult to compare directly the definition of $p$ and $\tau_0$ with
the expected values for real disks. We use disk parameters obtained in earlier
works to give observational expectations and also to help design future
observations.
  
  For disks, the fraction of the obscured to the total projected area, $p$
(Equation~\ref{def:p}), varies between $0$ when there is no obscuration and 0.5 for
a disk which obscures half of the stellar surface.
However, if the infrared emission region is not a disk but a shell around the
star \citep{rea05} $p$ will be always $1$. Hence, this work provides an
independent method to test the disk hypothesis.

  In addition to $p$, we can also determine $\tau_0$. Estimates for the expected
values are more uncertain, but we can get a rough idea by using some mean values
for the dust and disk properties. \cite{kru03} gives $2.5$~g/cm$^3$ as a typical
value for the interstellar dust and we assume it as a good order of magnitude
value for the dust in the disk. \cite{jur07b} constrain the disk mass of GD~362
between $10^{18}$ and $10^{24}$~g. Using typical disk sizes of $10~R_{wd}$ and
$100~R_{wd}$ for the inner and outer disk radii, respectively, we get a range
of $\tau_0$ from $10^{-4}$ to somewhat greater than $1$, depending on the disk
mass and the type of dust \citep{dor95}. Therefore the parameters used in Figure
\ref{pyrmg70} are realistic.

  The disk inclination angle can be inferred from the presence of an flux excess
due to scattering into the line of sight. Figure \ref{pyrmg70sca} shows this for
inclinations of $0^\circ$ and $60^\circ$. Flux excess in near-UV has already
been detected by \cite{gan2006} in SDSS~1228+1040 and could be caused by light
scattering. For larger inclinations there is a flux deficiency due to
absorption and scattering out the line of sight, as shown in
Figure \ref{pyrmg70} computed for $i=85^\circ$. The dividing line between
the first or the second case is $\sim$$80^\circ$.

\section{Conclusions}
\label{sec:conc}

In this work, we introduce a new way of looking at the cause of IR excess in
white dwarf stars.
By looking in the near-UV and optical instead of IR we add a new constraint to
test the disk hypothesis.

One important distinction of our method, is the fact that the presence of disks
would cause flux deficiencies in some systems and flux excess in others. We also
point out that shells would only introduce flux deficiencies effects, and these
effects would be detectable in all shells. If we find flux deficiencies in every
star we observe this would strongly indicate the presence of shells rather than
disks. Flux deficiencies in only some objects and flux excess in others
corroborate the idea of a disk.

If we are convinced disk models are more adequate, detailed
comparisons between disk models and data will provide disk 
mass \citep{jur07b}, composition, optical depth and inclination
relative to the line of sight.

\begin{acknowledgements}
  The authors acknowledge financial support from CNPq-MCT/Brazil. We thank
Paola D'Alessio, Detlev Koester, and Don Winget for helpful discussions,
William Reach for providing data, and Shashi Kanbur for reading the manuscript.
We are also grateful to Nikolai Voshchinnikov for his Mie-Theory program.
Finally, we thank the anonymous referee for helpful comments on paper.
\end{acknowledgements}

\end{document}